\documentstyle[epsfig]{mn2e}
\begin{document}
\input BoxedEPS.tex
\newcommand{\aip}{{\small ${\cal AIPS}$}}
\newcommand{\gtsim}{\mbox{{\raisebox{-0.4ex}{$\stackrel{>}{{\scriptstyle\sim}}
$}}}}
\newcommand{\ltsim}{\mbox{{\raisebox{-0.4ex}{$\stackrel{<}{{\scriptstyle\sim}}
$}}}}
\newcommand{\s}{$\stackrel{\rm s}{.}$}
\newcommand{\h}{$^{\rm h}$}
\newcommand{\m}{$^{\rm m}$}
\newcommand{\pp}{$\stackrel{\prime\prime}{.}$}
\newcommand{\de}{$^{\circ}$}
\newcommand{\p}{$^{\prime}$}
\newcommand{\arc}{$^{\prime\prime}$}
\newcommand{\marc}{^{\prime\prime}}
\newcommand{\rs}{{\em $r_s$}}
\newcommand{\DPM}{{\em DPM}}
\newcommand{\alf}{{\displaystyle\biggl({\nu_{\rm h} \over \nu_{\rm l}}\biggr)^{\alpha}} }

\newcommand{\figstart}[1]
    { \begin{figure}[htb]
      \begin{picture}(0,#1) }
\newcommand{\figend}[4]
    { \end{picture}
      \special{#1}
      \caption[#2]{#3}
      \label{#4}
      \end{figure} }
\newcommand{\fig}[5]
    { \figstart{#1}
      \figend{#2}{#3}{#4}{#5} }
\newcommand{\bHS}{\beta_{\mbox{\scriptsize HS}}}
\newcommand{\bBF}{\beta_{\mbox{\scriptsize BF}}}
\newcommand{\nT}{\nu_{\mbox{\scriptsize T}}}
\newcommand{\et}{E_{\mbox{\scriptsize T}}}
\newcommand{\nTn}{\nu_{\mbox{\scriptsize Tn}}}
\newcommand{\nTf}{\nu_{\mbox{\scriptsize Tf}}}
\newcommand{\tn}{\tau_{x\mbox{\scriptsize n}}}
\newcommand{\tf}{\tau_{x\mbox{\scriptsize f}}}
\newcommand{\xn}{x_{\mbox{\scriptsize n}}}
\newcommand{\xf}{x_{\mbox{\scriptsize f}}}
\newcommand{\yn}{y_{\mbox{\scriptsize n}}}
\newcommand{\yf}{y_{\mbox{\scriptsize f}}}
\newcommand{\lln}{l_{\mbox{\scriptsize n}}}
\newcommand{\llf}{l_{\mbox{\scriptsize f}}}
\newcommand{\Dn}{f(\Delta_{\mbox{\scriptsize n}})}
\newcommand{\Df}{f(\Delta_{\mbox{\scriptsize f}})}
\newcommand{\B}{\mbox{$B$}}
\newcommand{\Bo}{\mbox{$B$}_{0}}

\SetEPSFDirectory{/scratch/sbgs/figures/hst/}
\SetRokickiEPSFSpecial
\HideDisplacementBoxes

\title[Modelling JWST mid-infrared counts]{Modelling JWST mid-infrared counts: excellent consistency with models derived for IRAS, ISO and Spitzer}
\author[Rowan-Robinson M.]{Michael Rowan-Robinson\\
Astrophysics Group, Blackett Laboratory, Imperial College of Science 
Technology and Medicine, Prince Consort Road,\\ 
London SW7 2AZ}
\maketitle
\begin{abstract}
Models derived in 2009 to fit mid-infrared (8-24 micron) source counts from the IRAS, ISO and Spitzer missions, provide an excellent fit 
to deep counts with JWST, demonstrating that the evolution of dusty star-forming galaxies is well understood. 
The evolution of dust in galaxies at high 
redshifts is discussed and a simple prescription is proposed to model this.  This allows more realistic models for source-counts at submillimetre 
wavelength.  A reasonable fit to 250, 500, 850 and 1100 micron counts is obtained.
This paper therefore draws together the IRAS, ISO, {\it Spitzer}, {\it Akari}, {\it Herschel}, submillimetre ground-based, and JWST 
surveys into a single picture.

\end{abstract}


\section{Introduction}

The advent of deep JWST source-counts at mid-infrared wavelengths (Ling et al 2022, Wu et al 2023) poses the question: how do these 
relate to the counts made with IRAS, ISO, {\it Spitzer} and {\it Akari}?  

Source-counts at infrared and submillimetre wavelengths, combined with the spectrum of the 
infrared background, give us important constraints on the star-formation history of the universe.
First indications of strong evolution in the properties of star-forming galaxies came from IRAS 
60 $\mu$m counts (Hacking and Houck 1987, Lonsdale et al 1990, Rowan-Robinson et al 1991,
Franceschini et al 1991, 1994, Pearson and Rowan-Robinson 1996).

ISO gave us deep counts at 15 $\mu$m, providing strong evidence for evolution
(Oliver et al 1997,  Rowan-Robinson et al 1997,  Guiderdoni et al 1998, Aussel et al 1999, 
Elbaz et al 1999, 2002, Serjeant et al 2000, Gruppioni et al 2002, Lagache et al 2003) and useful 
counts at 90 and 175 $\mu$m
(Kawara et al 1998, Dole et al 2001, Efstathiou et al 2000a, Heraudeau et al 2004).

850 $\mu$m counts with SCUBA and SCUBA2 on JCMT (Smail et al 1997, Hughes et al 1998, 
Eales et al 1999, 2000, Barger et al 1999, Blain et al 1999, Fox et al 2002, Scott et al 2002,
Smail et al 2002, Cowie et al 2002, Webb et al 2003, Borys et al 2004, Coppin et al 2006, Chen et al 2013, Hsu et al 2016, 
Geach et al 2017, Simpson et al 2019, Shim et ak 2020, Hyun et al 2023) and with ALMA (Kamin et al 2013, Oteo et al 2016, 
Stach et al 2018, Simpson et al 2020, Gomez-Guijarro et al 2022), and counts at 1100 $\mu$m (Perera et al 2008, 
Scott et al 2012, with Aztec on JCMT; and Fujimoto et al 2016, Hatsukade et al 2018, Gomez-Guijarro et al 2022 with ALMA) have given
 important insights into the role of cool dust and also strong constraints on the high-redshift evolution of 
hyperluminous infrared galaxies. 

With {\it Spitzer} we also have deep counts at 8, 24, 70, 160 $\mu$m (Fazio et al 2004, 
Chary et al 2004, Marleau et al 2004, Papovich et al 2004, Dole et al 2004, Le Floch et al 2005, 
Frayer et al 2006, Shupe et al 2008, Clements et al 2011), with 24 $\mu$m providing 
an especially complete picture of the contribution of individual sources to the  infrared 
background.  While some pre-{\it Spitzer} models were quite successful at 70 and 160 $\mu$m,
none captured the full details of the 24 $\mu$m counts.  Lagache et al (2004) provided early revised models
in the light of the Spitzer data.  Franceschini et al (2010) also successfully modelled Spitzer counts data, but with an unphysical model with two power-laws separated by a discontinuity.

{\it Herschel} provided counts at 250, 350 and 500 $\mu$m, which proved challenging for pre-existing counts
models. Gruppioni et al (2011) successfully modelled the counts with 5 spectral types, including 3 AGN types, and both luminosity 
and density evolution.  Bethermin et al (2012) also successfully fitted these counts, but with an unphysical evolution model 
with three different power-laws separated by two discontinuities.  Cai et al (2015) modelled counts from 15 to 1380 $\mu$m 
with a model describing the evolution of photo-spheroidal galaxies and their associated AGN.  Bisigello et al modelled counts 
from 11 to 850 $\mu$m with 8 spectral types
and broken power-law luminosity and density evolution.  Bethermin et al (2017) discuss the impact of clustering and angular
resolution on far infrared and submillimtere counts.  Davidge et al (2017) gave counts at 3.2-24 $\mu$m from {\it Akari}.

Rowan-Robinson (2001, 2009) modelled infrared source-counts and background in terms of four
types of infrared galaxy: quiescent galaxies in which the infrared radiation (infrared 'cirrus')
is emission from interstellar dust illuminated by the general stellar radiation field; starbursts
for which M82 is the prototype; more extreme, higher optical depth starbursts with Arp 220 as 
prototype; and AGN dust tori.  In Rowan-Robinson (2001) the same evolution history, intended to reflect the global star-formation
history, was used for each galaxy type.  This model was consistent with counts, luminosity 
functions and colour-luminosity relations from IRAS, counts from 
ISO and  available integral 850 $\mu$m counts.  However the advent of deep source-count data from
{\it Spitzer} showed that this model, along with other pre-{\it Spitzer} models, failed, especially 
at 24 $\mu$m.  Rowan-Robinson (2009) showed how this model had to be modified to achieve consistency
with {\it Spitzer} data, specifically by allowing different evolution histories for each spectral component.

In this paper I show how my 2009 model provides an excellent fit to the deep JWST mid-infrared source-counts, and show 
that with an improved model for dust evolution at high redshift, counts at 250-1100 $\mu$m can also be fitted with the 
same evolutionary model, while at the same time maintaining consistency with the star-formation rate density history 
and the integrated background spectrum.

A cosmological model with $\Lambda$ = 0.7, $h_0$=0.72 has been used throughout.

\section{Methodology}

The philosophy of Rowan-Robinson (2001, 2009) was to find a simple analytic form for the 
evolutionary function, without discontinuities and involving the minimum number of parameters.
Both papers used the same four basic infrared templates: (1) quiescent galaxies, radiating in the infrared through 
emission by interstellar dust of absorbed starlight (' infrared cirrus'), (2) starbursts, with prototype 
M82, (3) extreme starbursts with much higher dust optical depth, with prototype Arp 220, (4) AGN
dust tori.  These templates (illustrated in Fig 3 of Rowan-Robinson (2001)), which have been derived through radiative 
transfer calculations (Efstathiou and Rowan-Robinson 1995, 2003, Rowan-Robinson 1995, Efstathiou et al 2000b), proved 
extremely effective in modeling the spectral
energy distributions (SEDs) of infrared galaxies in the ISO-ELAIS (Rowan-Robinson et al 2004),
 {\it Spitzer}-SWIRE (Rowan-Robinson et al 2005), and SHADES (Clements et al 2008) surveys.
For the {\it Herschel} surveys, additional components (cool dust, young starbursts) were required 
(Rowan-Robinson 2010, 2014), but these have not been added to the counts models here.  
The luminosity functions for each 
component are derived as in Rowan-Robinson (2001), from the 60 $\mu$m luminosity function
via a mixture table which is a function of 60 $\mu$m luminosity (Table 2).  The 2009
luminosity functions were essentially unchanged from Rowan-Robinson (2001), except that the
Arp 220 (A220) component was assumed to be more significant in the luminosity range
$L_{60} = 10^9-10^{10} L_{\odot}$, in order to improve the fit to the counts at 850 $\mu$m.
Here some further adjustments to the mixture table at low luminosities were made to fit the star formation rate density versus
redshift (section 7). These adjustments have negligible effect on the source-counts.

\begin{table}
\caption{Evolutionary parameters for each component}
\begin{tabular}{lllll}
component & P& Q& $a_0$ & $\tau_{sf}$\\
 &&&& (Gyr)\\
&&&&\\
cirrus & 4.0 & 12.0 & 0.8 & 1.15\\
M82 & 2.7 & 18.0 & 0.994 & 0.77\\
A220 & 2.7 & 14.0 & 0.92 & 1.01\\
AGN & 2.7 & 13.0 & 0.94 & 1.06\\
\end{tabular}
\end{table}

\begin{table}
\caption{Proportion of different SED type at 60 $\mu$m as a function of $L_{60}$}
\begin{tabular}{lllll}
 & quiescent & M82 & A220 & AGN dust\\
$lg_{10} (L_{60}/L_{\odot})$ & & starburst & starburst & torus\\
&&&&\\
8.0 & 0.99 & 0.001 & 0.5e-10 & 0.009\\
9.0 & 0.99 & 0.0049 & 0.0049 & 0.0012\\
10.0 & 0.80 & 0.10 & 0.10 & 0.5e-4\\
11.0 & 0.10 & 0.45 & 0.45 & 0.19e-4\\
12.0 & 0.04 & 0.48 & 0.48 & 0.245e-6\\
13.0 & 0.1e-7 & 0.50 & 0.50 & 0.245e-8\\
14.0 & 0.1e-12 & 0.50 & 0.50 & 0.245e-10\\
\end{tabular} 
\end{table}

The main new feature of the 2009 work was that I allowed each component its own evolution 
function, since a single evolution function for all components was not capable of reproducing the
latest counts, especially at 24 $\mu$m.

In Rowan-Robinson (2001) I assumed pure luminosity evolution, with $L(z) = \phi(z) L(0)$, and

$\phi(z) = exp [ Q (t_0-t)/t_0 ] (t/t_0)^P$		(1)

where $t_0$ is the current epoch, the exponential factor is essentially the Bruzual and Charlot  
(1993) star-formation history
and the power law factor ensures that L(z) tends to zero as $t$ tends to 0.  Thus $t_0/Q$ has the meaning
of the exponential time-scale for star-formation.  The peak star-formation rate in this model
occurs at $t/t_0 = P/Q$.  

However the 24 $\mu$m counts show an extended Euclidean behaviour at bright fluxes and in the 2009 paper
I therefore allowed for the possibility that the evolution function tends asymptotically to a
constant value at late times.  I also allow the epoch where the evolution function tends to zero
at early times to be $t_f = t(z_f)$ rather than zero, where $z_f$ is to be determined from the
fits to the counts.  This option was found to be important at submillimetre wavelengths.  Thus
I assumed

$\phi(z) = [ a_0 + (1-a_0) exp  Q (1-y )]  (y^P -(y_f)^P)$                 (2)

where $y = (t/t_0)$, $y_f = t(z_f)/t_0$.

Rowan-Robinson (2009) described the strategy for determining the parameters P, Q, $a_0$  for each component.
$z_f$ represents the epoch at which the dust abundance in star-forming regions becomes high enough
for those regions to be optically thick.  Rowan-Robinson (2009) argued that $z_f$ could be as low as
4, given that it takes a billion years or so for Population II red giants to evolve and start 
outputting dust to the interstellar medium. However $\it Herschel$ has provided abundant evidence for
dusty starbursts at z =4-6 and beyond. JWST has given plenty of evidence of galaxies at $z \sim 10$, but not many
at redshifts higher than that, so we adopt $z_f = 10$ in this paper.


The predicted counts at 850 (and 1100) $\mu$m depend sensitively on the evolutionary 
parameters for A220 starburst component, but these parameters are essentially undetermined
at 24 $\mu$m.  The evolutionary parameters for the A220 component
were adjusted in Rowan-Robinson (2009) to give an equal contribution of A220 and M82 starbursts at S(850) = 8 mJy,
since SED modelling of SHADES sources (Clements et al 2008) suggests both templates are important at
these fluxes.  This also required an increase in the luminosity function for A220 sources 
at lower luminosities.  Evidence that submillimetre sources
do not constitute a monolithic, Arp-220-like, population was given by Menendez-Delmestre
et al (2007), Ibar et al (2008 and Pope et al (2008).

Adopted values of P, Q, $a_0$ for each of the four components are given in Table 1.
The star-formation time-scale, $\tau_{sf}$, is shown for an assumed $t_0$ =13.8 Gyr (Planck team 2106).
I should emphasize that the assumed evolution is applied to all infrared luminosities, and
I find no need to postulate very strong evolution only of luminous galaxies as has been 
proposed in many source-count models in the literature.

The assumption of pure luminosity evolution is perhaps questionable, since we know that
galaxies grow by mergers, so that we expect some steepening of the low luminosity end of the 
luminosity function with increasing redshift (see section 6).

\begin{figure*}
\epsfig{file=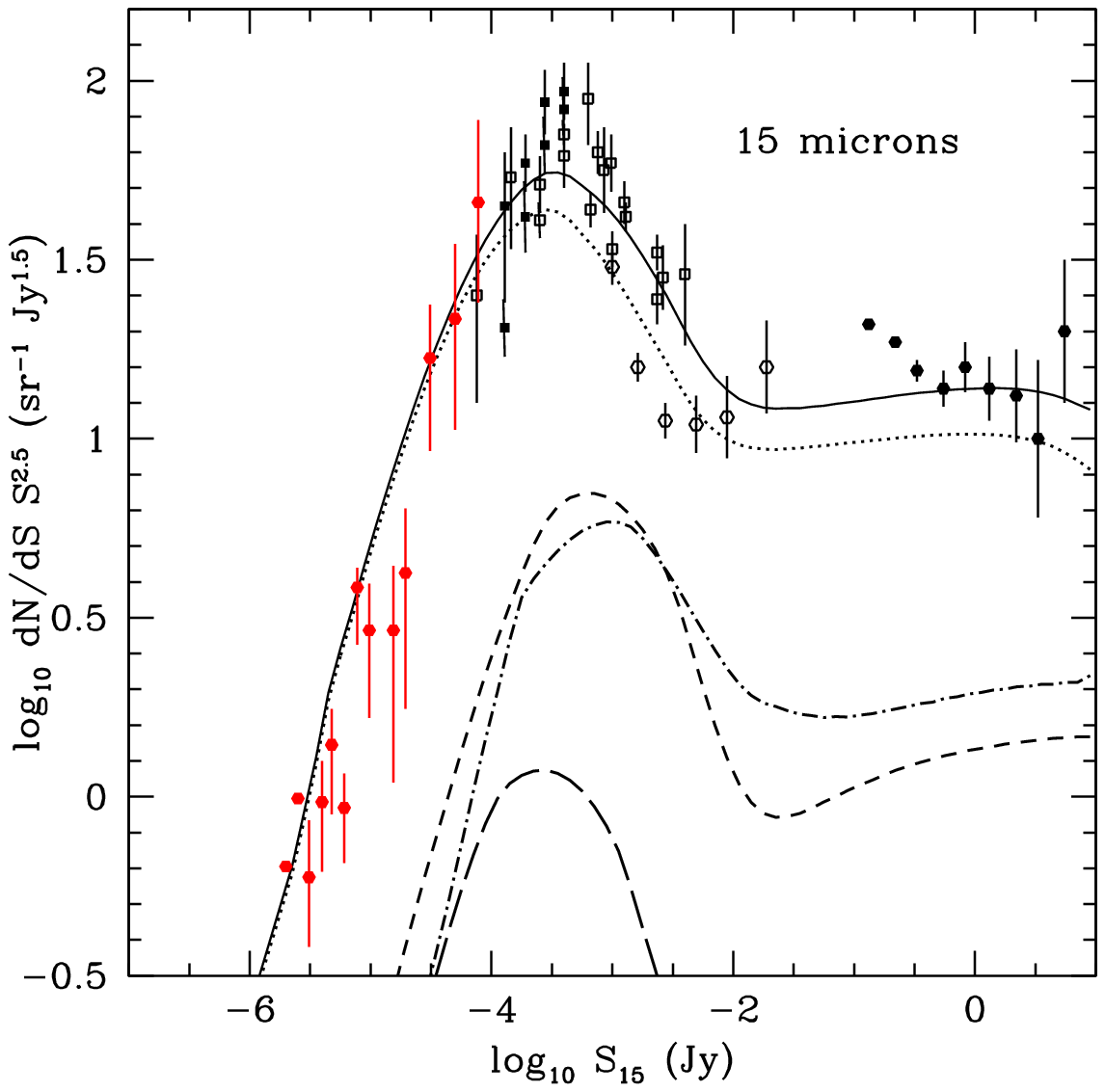,angle=0,width=14cm}
\caption{Euclidean normalised differential counts at  15  $\mu$m.
Solid curve: total counts (black: $z_f$ = 10) ; dotted curve: cirrus; short-dashed curve: M82;
long-dashed curve: A220; dash-dotted curve: and AGN dust tori.
ISO data from Elbaz et al 1999 (open squares), Aussel et al 1999 (filled squares), Gruppioni et al 2002
(open hexagrams); interpolated IRAS data (black filled hexagrams) from Verma (private communication); Akari data from Davidge et al 2015
(filled black triangles); JWST data from Wu et al 2023 (filled red hexagrams).
Solid curve: total counts; dotted curve: cirrus; short-dashed curve: M82;
long-dashed curve: A220; dash-dotted curve: AGN dust tori. 
}
\end{figure*}

\begin{figure*}
\epsfig{file=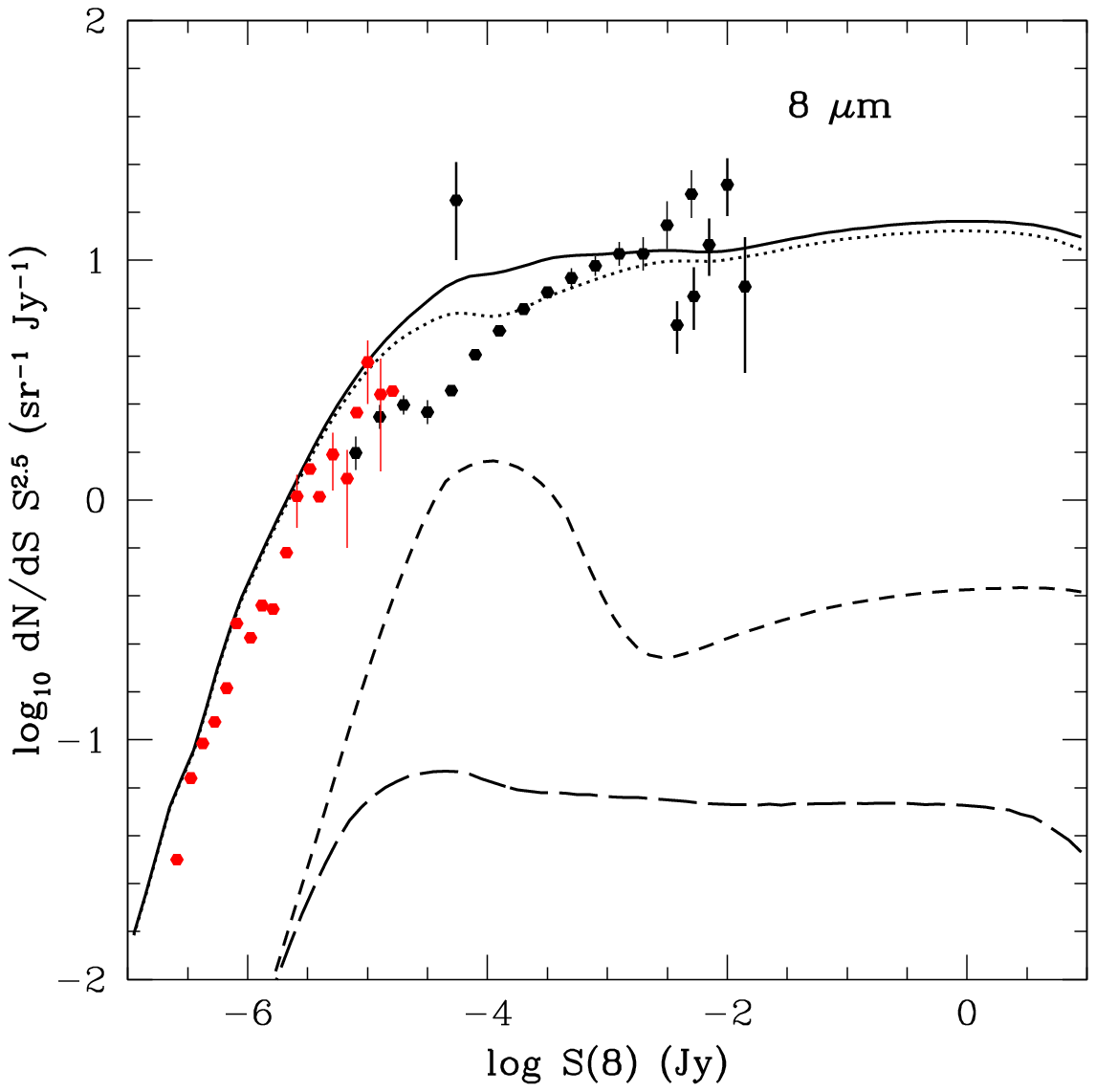,angle=0,width=7cm}
\epsfig{file=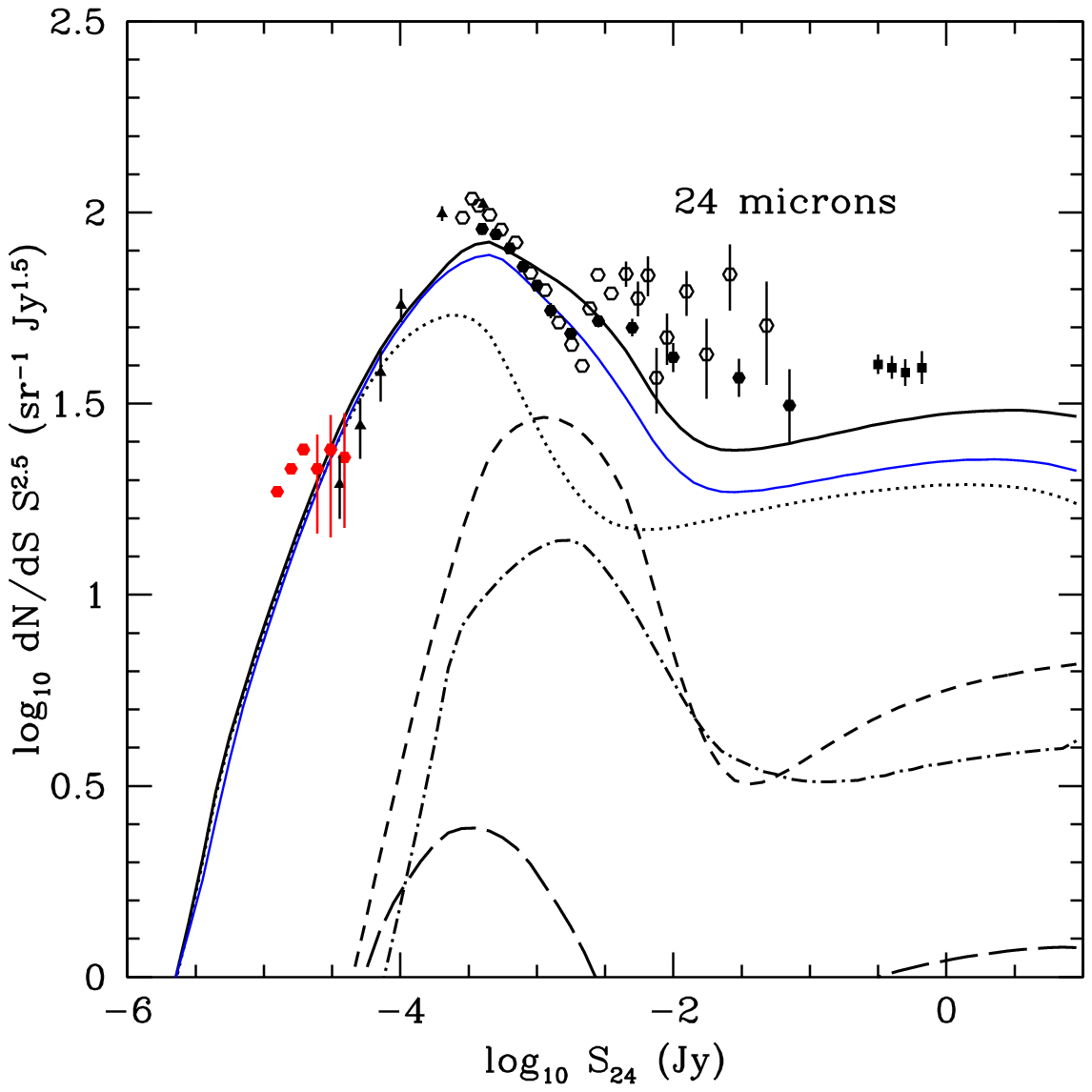,angle=0,width=7cm}
\caption{
Euclidean normalised differential counts at 8 (LH) (Spitzer data from Fazio et al 2004, black filled hexagrams; JWST from Wu et al 2023, 
red filled hexagrams) and 24 (RH) $\mu$m (IRAS data from Verma (private communication), filled squares.  Spitzer data from Shupe et al 2008, 
filled hexagrams; Papovich et al 2004, filled triangles; and Clements et al 2011, black filled squares.
JWST data from Wu et al 2023, filled red hexagrams.) Predicted loci as in Fg 1.  Blue locus is predicted 21 $\mu$m counts.
}
\end{figure*}

\section{Fits to JWST mid-infrared counts}

Here I discuss the fits to the observed JWST counts at 7.7, 15 and 21 $\mu$m (Wu et al 2023).  

Fitting the ISO and {\it Spitzer} counts at 15 and 24 $\mu$m played an important part in determining the 
evolutionary parameters of the 2009 model.  Figures 1 and 2 show the fits to the counts
at 7.7, 15 and 24 $\mu$m.  IRAS, ISO, {\it Spitzer} and {\it Akari} data at 8, 15 and 24/25 $\mu$m are included in these plots.  To reduce clutter 
on these and subsequent count plots data points are excluded if they are $< 3-\sigma$ measurements.
What seems remarkable is that this model now provides an
almost perfect fit to the JWST counts at these wavelengths, even thought the JWST counts are 40 times deeper than
the Spitzer counts at 8 and 15 $\mu$m, reflecting the fact that the mirror diameter of JWST is 8 times greater
than that of {\it Spitzer}.  The dominant contribution at faint fluxes is from quiescent galaxies, since these 
dominate at lower infrared luminosities.  Note that at fluxes $<$ 0.1 mJy counts at 21 $\mu$m are indistinguishable
from those at 24 $\mu$m.

The fit at intermediate fluxes at 24 and 15 $\mu$m (0.1-10 mJy) can be improved, without losing the fit at faint fluxes or at 
8 $\mu$m by a small change to the evolutionary parameters for the quiescent component from (Q, $a_0$) = (10, 0.6) to (12, 0.8) and this
change has been adopted throughout this paper.

\begin{figure*}
\epsfig{file=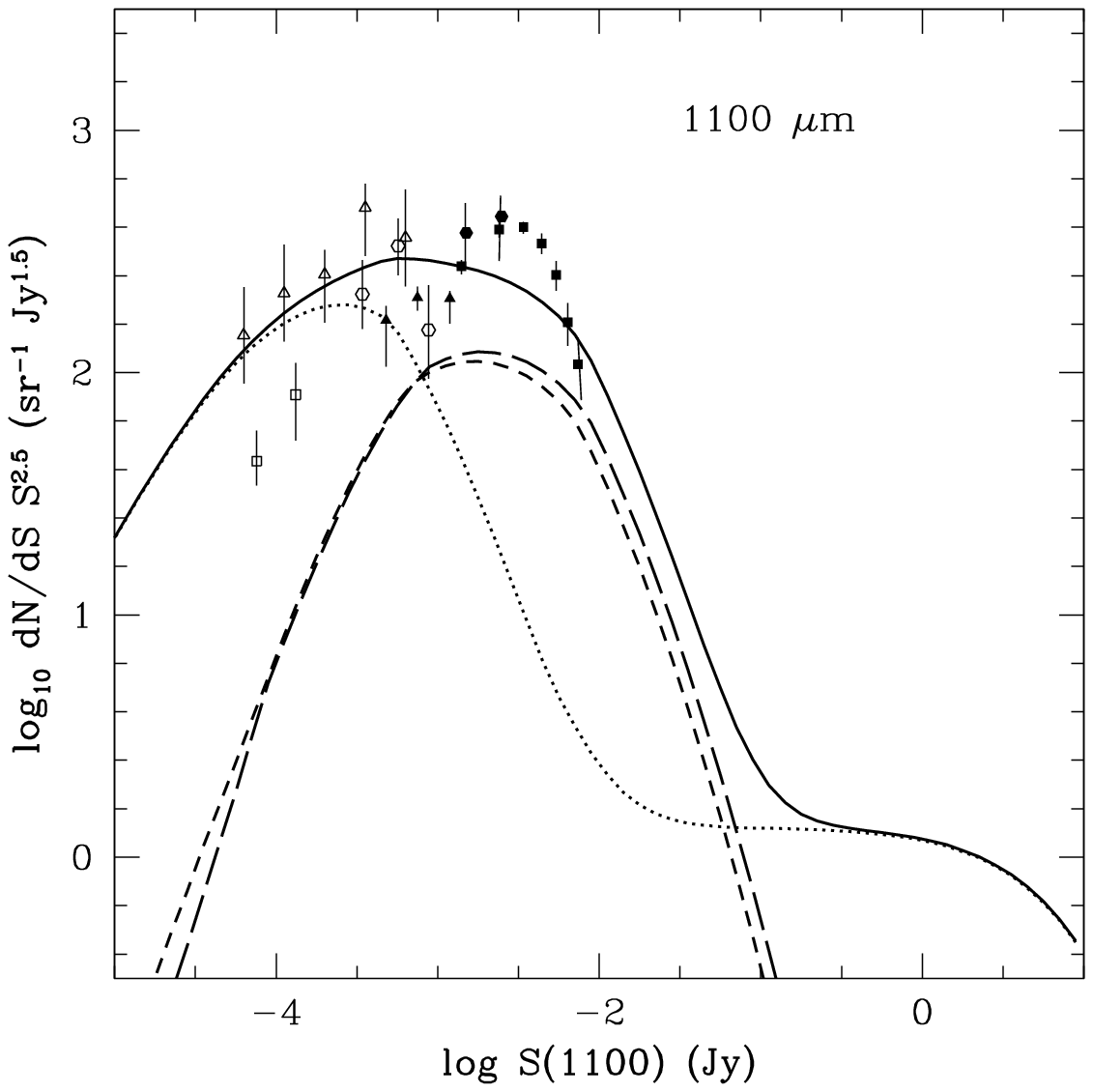,angle=0,width=7cm}
\epsfig{file=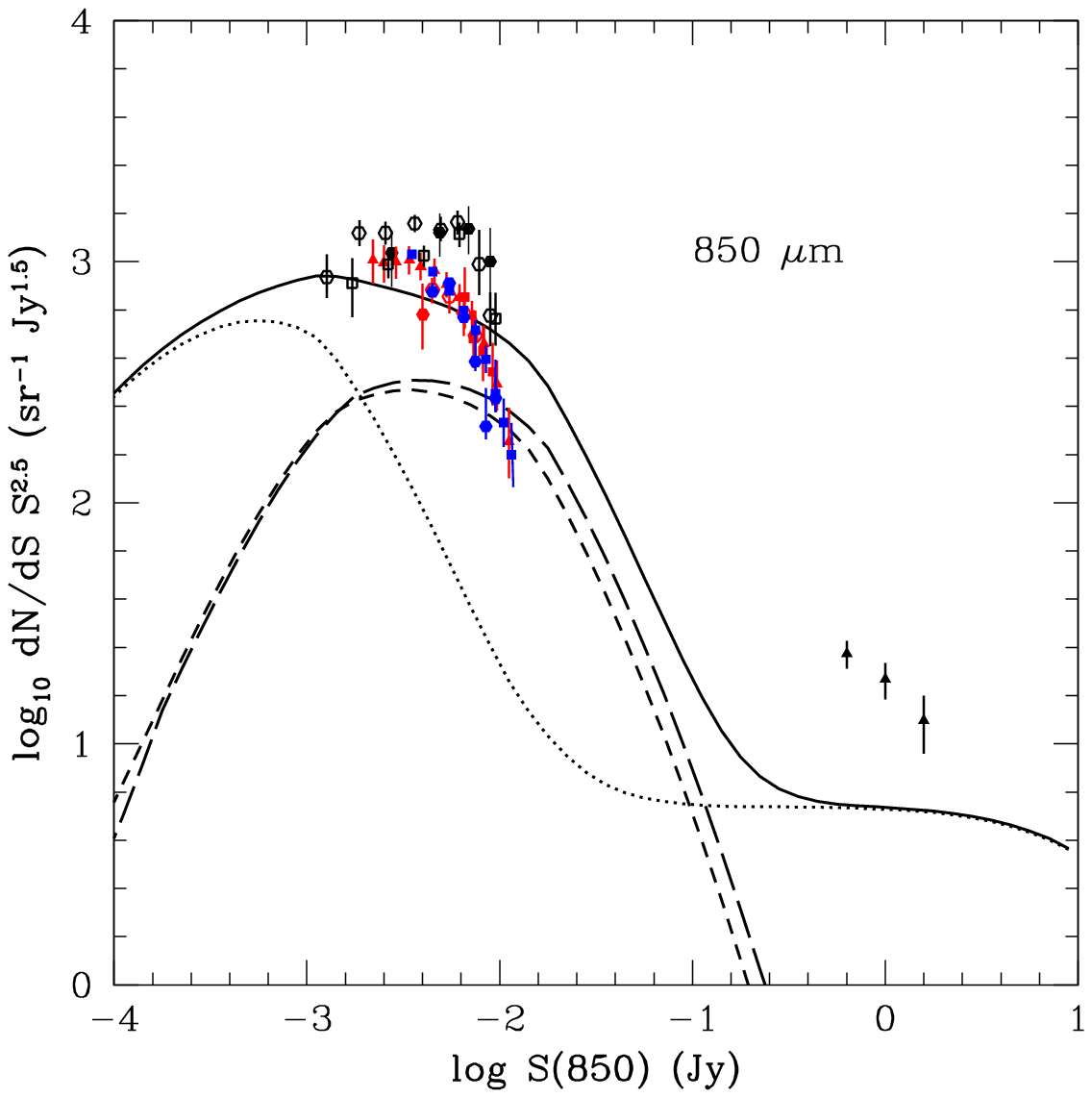,angle=0,width=7cm}
\epsfig{file=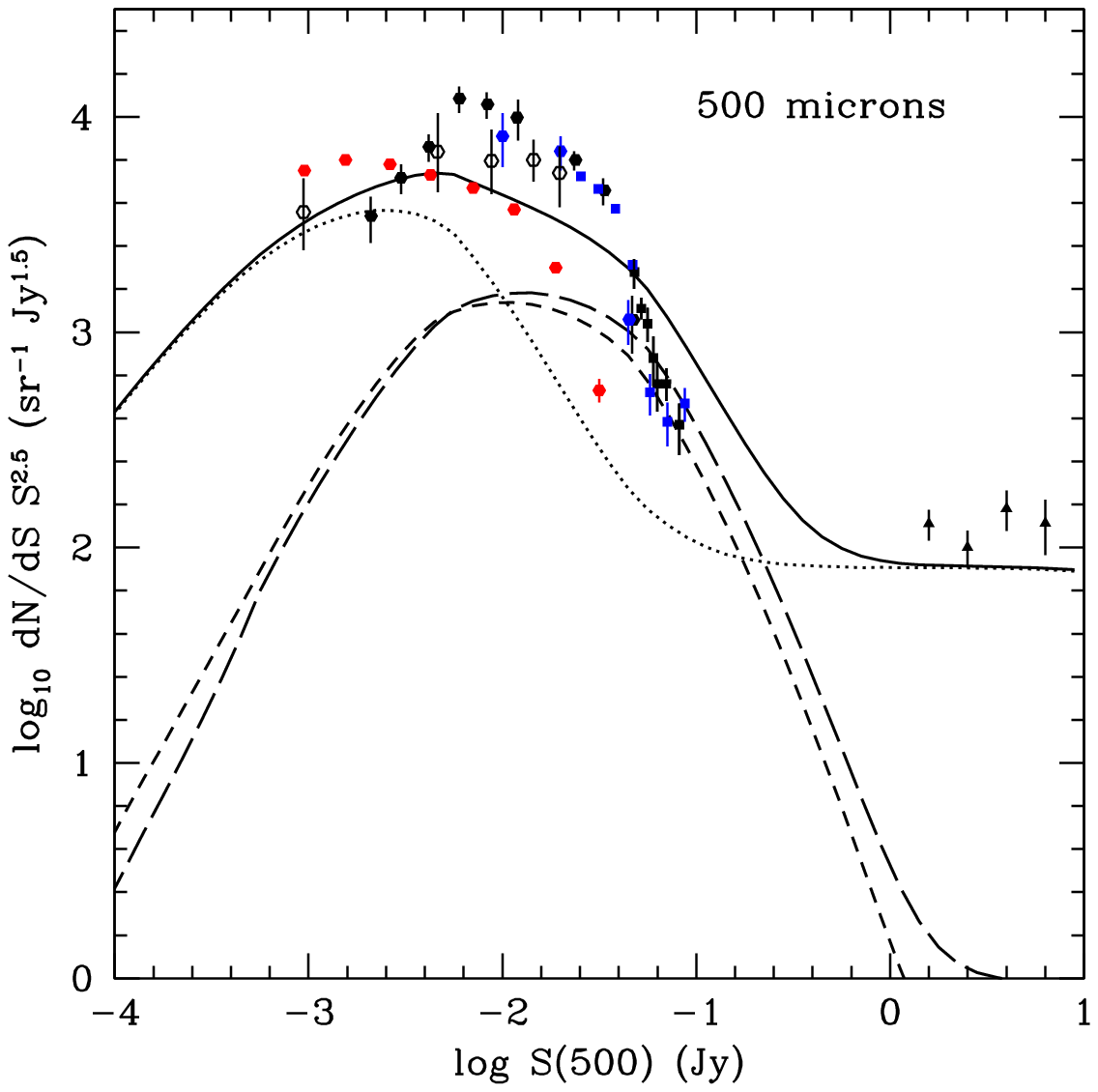,angle=0,width=7cm}
\epsfig{file=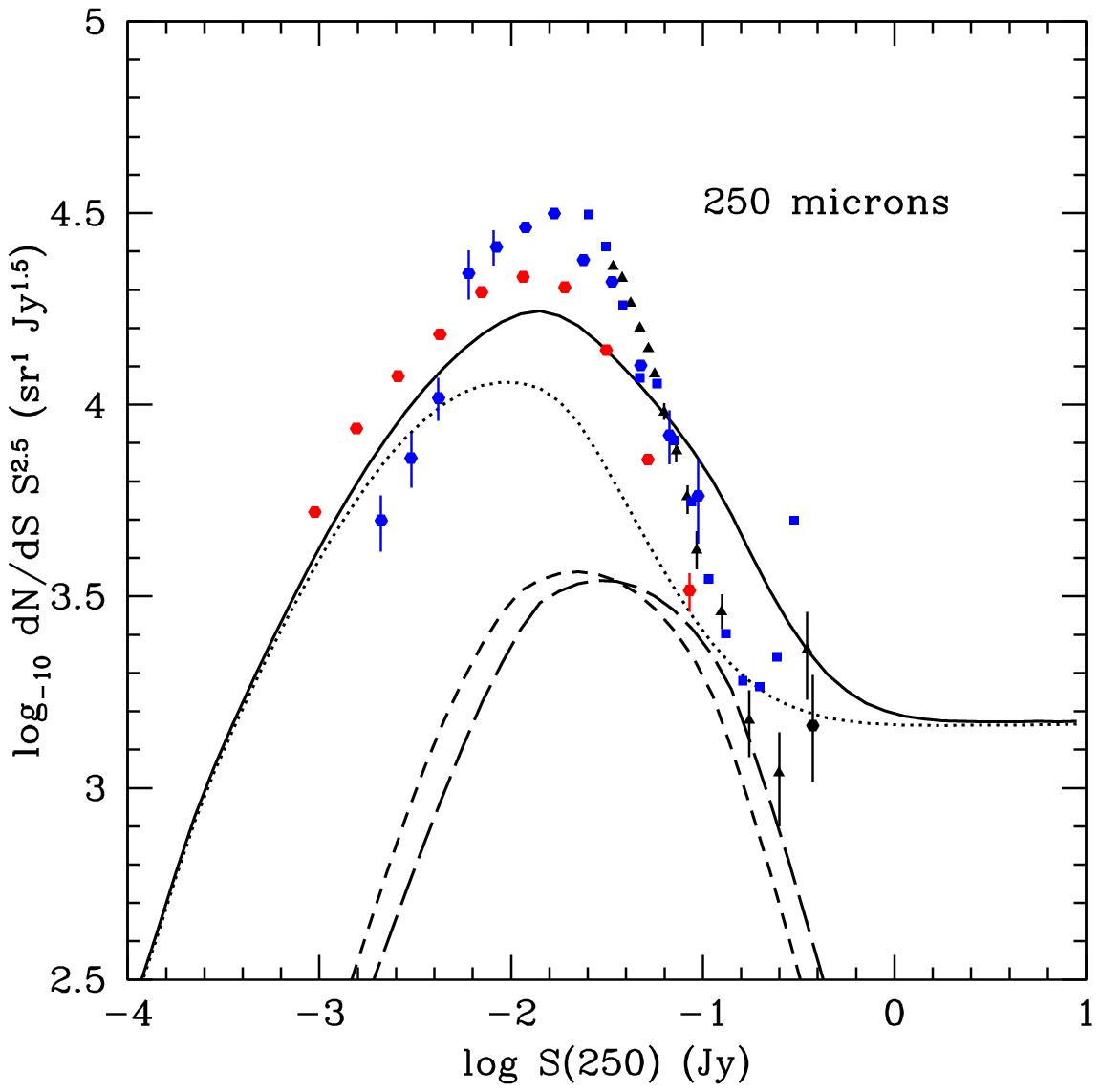,angle=0,width=7cm}
\caption{Euclidean normalised differential counts at 1100 (data from Perera et al 2008, filled black hexagons; Scott et al 2012, 
filled black squares; Fujimoto et al 2016, open black triangles; Hatsukade et al 2018, open black hexagrams; Gomez et al 2022, 
filled black triangles), 850 (data from  Coppin et al 2006, filled black hexagrams; Chen et al 2013, 
black open squares; Planck team 2013, filled black triangles; Hsu et al 2016, black open hexagrams; Geach et all 2017, 
blue filled squares; Stach et al 2018, blue filled
hexagrams; Simpson et al 2019, red filled triangles, Simpson et al 2020, red filled squares; Shim et al 2020, red open hexagrams;
Hyun et al 2023, red filled hexagrams.), 500 (data from 
Clements et al 2010, black filled triangles; Glenn et al 2010, blue filled hexagrams; Bethermin et al 2012, black filled hexagrams; 
Planck team 2013, black filled triangles; Valiante et al 2016, blue filled triangles; Hsu et al 2016, black open hexagrams;
Wang et al 2019, red filled hexagrams.) and 250 $\mu$m (data from Oliver et al 2010, black filled hexagrams; 
Clements et al 2010, black filled squares; Glenn et al 2010, blue filled hexagrams; Bethermin et al 2012, blue filled hexagrams; 
Valiante et al 2016, blue filled triangles; Wang et al 2019, red filled hexagrams.)
Solid curve: total counts; dotted curve: cirrus; short-dashed curve: M82;
long-dashed curve: A220; dash-dotted curve: AGN dust tori. 
}
\end{figure*}

\section{Effect of dust evolution on SEDs and sub-millimetre source counts}

In order to predict source-counts at submillimeter wavelengths, it is necessary to consider the evolution of the
dust mass in galaxies with redshift. This was modelled in Rowan-Robinson (2009) with a crude cut-off of the evolution at
$z_f \sim 4$, to represent the onset of dust formation from red giant stars. However it has subsequently become clear
that there are plenty of dusty infrared galaxies at z = 4-6, at least (Wang et al 2007, Younger et al 2007, Schinnerer et al 2008,
Rowan-Robinson et al 2016, 2019). In reality dust from supernovae will start to 
appear in galaxies as soon as the first massive stars form, and dust from red giant stars will start to be present
from about 0.5 billion years later. The details of dust evolution from z=10 to z=4 are not yet known (see Maiolino and Mannucci 2019
for a review of our knowledge of dust), but the evolution of metallicity with redshift has been discussed by Fynbo et al
(2006) and Wuyts et al (2016).  A reasonable approximation is that:
\\
	dust mass $\propto$ metallicity $\propto$ $(1+z)^{-Q_D}$, with $Q_D \sim 0.5$.
\\
At wavelengths $>$ 50 $\mu$m, the emission from dust in galaxies is essentially optically thin, so the
brightness will scale with the dust mass.  Thus instead of a dust cutoff at z $\sim$ 4, I propose that 
the submillimetre emission from galaxies scales as $(1+z)^{-Q_D}$ for $\lambda_{em} > 50 \mu$m, with $Q_D$ = 0.5.  Figure 3
shows counts at 250, 500, 850 and 1250 $\mu$m with this assumption.  No account is taken of lensing, which would increase the counts 
at brighter fluxes. The fits to the data are reasonable, particularly taking into account the quite wide spread in the counts
estimates, and the possible impact of resolution effects at the bright end (Bethermin et al 2017).  Good fits to these submillimetre data have been
given by Gruppioni et al (2011), Cai et al (2013), Bethermin et al (2017) and Bisigello et al (2021).


\begin{figure*}
\includegraphics[width=14cm]{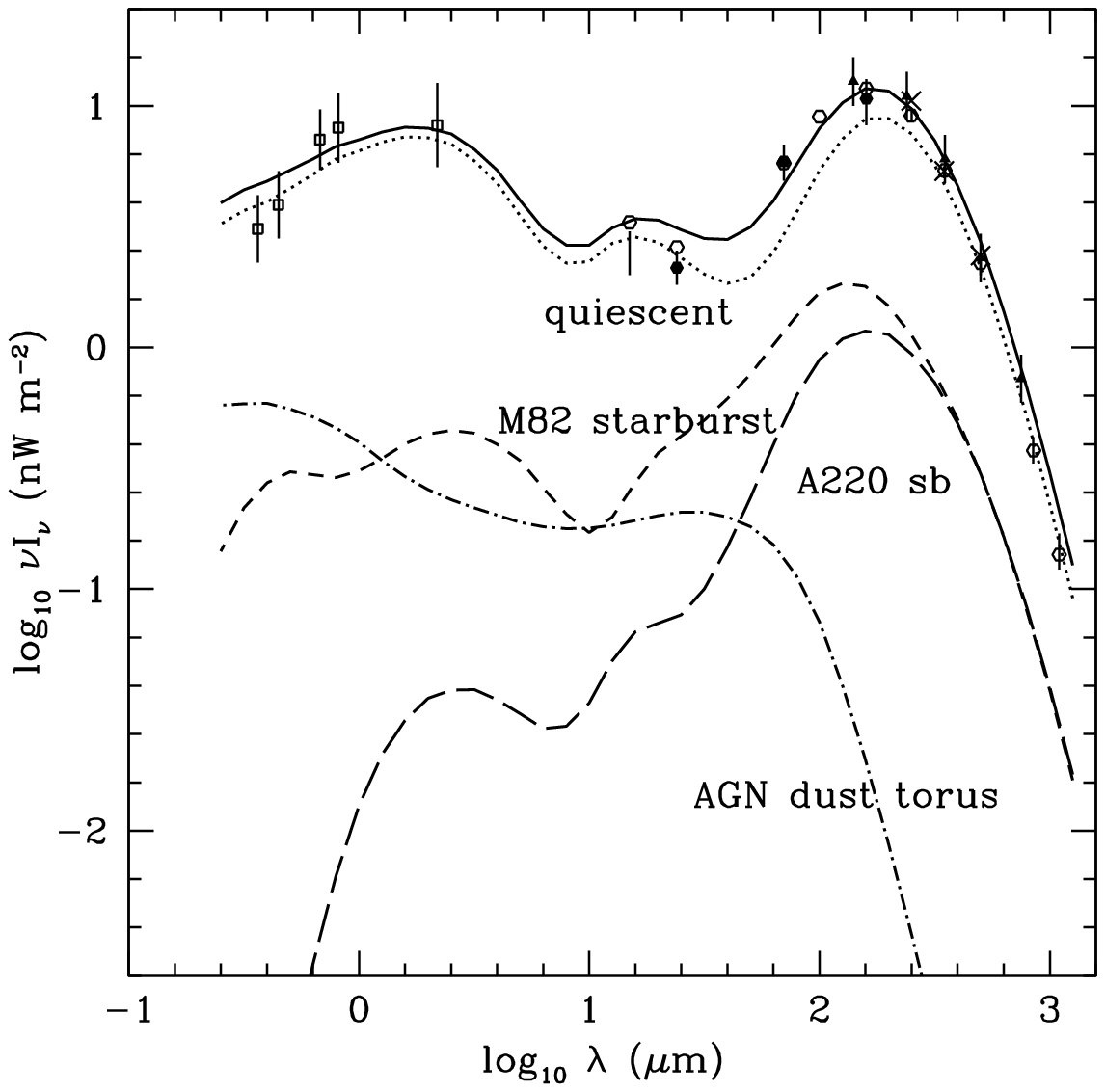}
\caption{
Background spectrum from 0.25 to 1250 $\mu$m. Data from Fixsen et al (1998, open triangles),
Dole et al (2006, filled circles), Serjeant et al (2000, vertical bar), Pozzerri et al (1998, open squares), Bethermin et al (2012, open circles).
Solid black locus: predicted background for model with QSED=0.5 and $z_f$=10. Other loci: as in Fig 1.
}
\end{figure*}

\section{Integrated background radiation}

The detection of the infrared background with COBE (Puget et al 1996, Fixsen et al 1998, Hauser et al 1998, 
Lagache et al 1999) provided an important constraint on evolutionary models
(Guiderdoni et al 1997, 1998, Franceschini et al 1998, 2001, Dwek et al 1998, Blain et al 1999, Gispert et al 2000,
Dole et al 2001, Rowan-Robinson 2001,  Chary and Elbaz 2001, Elbaz et al 2002,  King and RR 2003,
Balland et al 2003, Xu et al 2003, Lagache et al 2003).  Dole et al (2006) used stacking analysis on deep
{\it Spitzer} counts at 24, 90 and 160 $\mu$m to estimate the integrated background radiation from sources
at these wavelengths.   Bethermin et al (2012) give estimates of the integrated background derived from deep counts at 
16-850 $\mu$m.  Figure 5 shows the predicted
background spectrum derived from our counts model.  The fit from mid infrared to submillimetre wavelengths is excellent.  
We also show the fit in the optical and near IR, which reflects the assumed optical extensions of the IR templates. 
The fit at these wavelengths is surprisingly good, considering the crudity of the modelling at these wavelengths.


The infrared background spectrum is sensitive to the details of dust and luminosity function evolution at high redshift.
A more detailed model of the evolution of dust at early times is beyond the scope of this paper, but it is of
interest to consider whether our luminosity function evolution could be modified to reflect the actual merger
history of galaxies. The main effect expected is that the luminosity function steepens at high redshifts to 
reflect those small galaxies that will end up being merged into larger galaxies (eg Mashian et al 2016) and a general increase in the
number density of galaxies with increasing redshift.  
%
%
%
%
Faint 8 and 15 $\mu$m counts are sensitive to the faint-end luminosity function, so this may not be straightforward.

\begin{figure*}
\includegraphics[width=14cm]{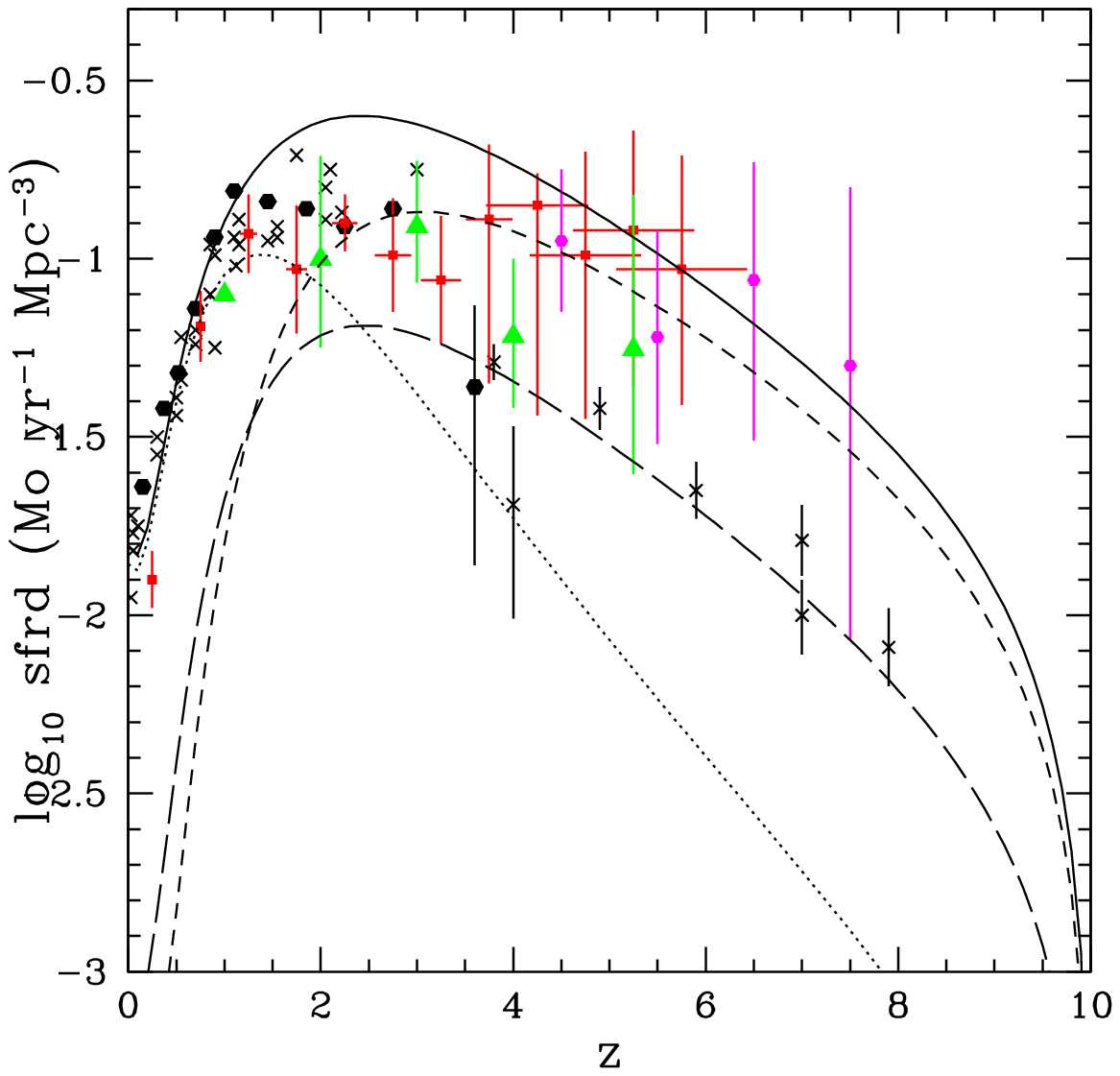}
\caption{
Star formation rate density as a function of redshift.  Crosses: optical and ultraviolet data summarised
by Madau and Dickinson (2014); black filled hexagons: far infrared data of Gruppioni et al (2013);
filled red squares: sum of starburst and quiescent galaxies from Herschel 500 $\mu$m sample (Rowan-Robinson et al 2016, 2019). 
Green triangles: Gruppioni et al 2020. Magenta hexagons: Kistler et al 2009 estimates for GRBs.
Solid locus: combined estimate from Rowan-Robinson (2009) source-counts model, as modified here (broken locus: M82 
starbursts, long-dashed locus: A220 starburst, dotted locus: quiescent galaxies).
}
\end{figure*}

\section{The history of star-formation}

Madau and Dickinson (2014) have reviewed optical and uv estimates of the star formation rate density as a function of redshift and 
Rowan-Robinson et al (2016) and Gruppioni (2020) have given estimates derived from {\it Herschel} data.
Kistler et al (2009) give a very interesting independent estimate derived from gamma-ray bursts.
Figure 6 shows these estimates compared with the predictions of our counts model.
This diagram does depend sensitively on the details of the infrared luminosity functions at low luminosities
and our assumptions are shown in Table 2.

\section{Discussion}

The model presented here, based on four basic infrared templates is obviously an oversimplification.  
We know that to understand local galaxies we need a range of cirrus components (Rowan-Robinson 1992, 
Efstathiou and Rowan-Robinson 2003).  
From the models of Efstathiou et al (2000) the SED of a starburst varies strongly with the age of the starburst
so the representation as two simple extremes, M82 and Arp220 starbursts, is oversimplified.
Finally we know that we need a range of dust torus models to fit ultraluminous
and hyperluminous galaxies (Efstathiou and Rowan-Robinson 1995, Rowan-Robinson 2000, Farrah et al 2003, 
Rowan-Robinson and Efstathiou 2009). 

However the four templates used here do capture the main features of the infrared galaxy population.  In fitting
the SEDs of ISO and {\it Spitzer} sources (Rowan-Robinson et al 2004, 2005, 2008), we allow individual sources
to include cirrus, M82 or Arp220 starburst, and an AGN dust torus, which gives a rich range of predicted
SEDs.  This works for SCUBA sources (Clements et al 2008) and for detailed IRS spectroscopy of
infrared galaxies (Farrah et al 2008).

The strongest evolution rate is found for M82 starbursts.  Arp220 starbursts peak at a slightly later redshift
(2 rather than 2.5) and at a slightly lower amplitude.   AGN dust tori, where the evolution reflects the 
accretion history, also peak at redshift 2, but show a less steep decline to the present than M82 or Arp220
starbursts, suggesting a gradual change in the relative efficiency with which gas is converted to stars
or accreted into a central black hole (cf Rowan-Robinson et al 2018). 

Quiescent galaxies show a much more modest evolution rate, peaking at redshift 1.  However even in relatively quiescent galaxies like our
own or M81, the star formation activity was greater at redshift 1 than at the present day.   


An interpretation of our assumed form of evolution is as follows: the power-law term represents the rapid increase of 
star-formation in the universe during the merger and interaction phase of galaxy assembly, while the exponential decline 
corresponds to the gradual exhaustion of gas in galaxies by star-formation.  Rowan-Robinson (2003) showed how 
this form of evolution fits into a simple closed box model for star-formation in galaxies.  

Future work, observational and theoretical, could improve the assumptions about dust evolution and about 
the evolution of the luminosity function at high redshift.  A very different approach is followed by Cowley et al
(2018), in which the GALFORM galaxy formation simulation is combined with a dust model to predict source-counts 
in the JWST mid-infrared wave-bands, with some success at faint fluxes (Wu et al 2023).

\section{Conclusions}

The counts model of Rowan-Robinson (2009) provides an excellent fit to the deep JWST counts at 7.7, 15 and 21 $\mu$m.
With a more realistic model for dust evolution with redshift, a reasonable fit is found for $\it Herschel$ and ground-based submillimetre
source-counts.  A good fit to the integrated background spectrum and the evolution of the star-formation rate density with redshift 
is obtained with an adjustment to the relative contribution of quiescent and starburst galaxies at low luminosities.

In this way the surveys with the IRAS, ISO, {\it Spitzer}, {\it Akari}, {\it Herschel} and JWST missions, and ground-based submillimetre
surveys with SCUBA, ALMA and other instruments, are brought into a single evolutionary picture.

\section{Acknowledgements}
I thank David Clements for helpful comments on an early draft.  I thank an anonymous referee for helpful comments, which have improved the paper.

\section{Data availability}
All data used in this paper are in the public domain.


\end{document}